# THE DISCRETE COSINE TRANSFORM OVER PRIME FINITE FIELDS

M.M. Campello de Souza, H.M. de Oliveira, R.M Campello de Souza, M. M. Vasconcelos

Federal University of Pernambuco - UFPE, Digital Signal Processing Group
C.P. 7800, 50711-970, Recife - PE, Brazil
E-mail: {hmo, marciam, ricardo}@ufpe.br, mmv@ee.ufpe.br

**Abstract** - This paper examines finite field trigonometry as a tool to construct trigonometric digital transforms. In particular, by using properties of the k-cosine function over GF($p$), the Finite Field Discrete Cosine Transform (FFDCT) is introduced. The FFDCT pair in GF($p$) is defined, having blocklengths that are divisors of ($p$+1)/2. A special case is the Mersenne FFDCT, defined when $p$ is a Mersenne prime. In this instance blocklengths that are powers of two are possible and radix-2 fast algorithms can be used to compute the transform.

**Key-words**: Finite Field Transforms, Mersenne primes, DCT.

## 1. Introduction

The manifold applications of discrete transforms defined over finite or infinite fields are very well known. The Discrete Fourier transform (DFT) has, since long, been playing a decisive role in Engineering. Another powerful discrete transform is the Discrete Cosine Transform (DCT), which became a standard in the field of image compression [1]. Such transforms, despite being discrete in the variable domain, have coefficients with amplitudes belonging to an infinite field. They can therefore be understood as some kind of "analog transforms" (something as Pulse Amplitude Modulation systems). In contrast, transforms defined over finite fields, are discrete in both variable and transform domain. Their coefficients are from a finite alphabet, so they can be understood as "Digital Transforms". They can possibly be attractive in the same extent as digital systems are, compared to analog systems. A Fourier analysis can also be applied for analyzing signals over a finite field.

A very rich transform related to the DFT is the Finite Field Fourier Transform (FFFT), introduced by Pollard in 1971 [2] and applied as a tool to perform discrete convolution using integer arithmetic. Since then several new applications of the FFFT have been found, not only in the fields of digital signal and image processing [3-5], but also in different contexts such as error control coding and cryptography [6-7].

The Finite Field Hartley Transform, which is a digital version of the Discrete Hartley Transform, has been recently introduced [8]. Its applications include the design of digital multiplex systems, multiple access systems and multilevel spread spectrum digital sequences [9-12].

Among the discrete transforms, the DCT is specially attractive for image processing, because it minimises the blocking artefact that results when the boundaries between subimages become visible. It is known that the information packing ability of the DCT is superior to that of the DFT and other transforms. Furthermore, the DCT is not data dependent like the Karhunen-Loève transform, the best of all linear transforms for energy compaction [13]. The DCT is now a tool in the international JPEG standard for image processing, but there is no equivalent transform over finite fields. A natural question that arises is whether it is possible to find a DCT representation over such fields. The first task towards such a new transform is to establish the equivalent of the cosine function over a finite structure.

Based on the trigonometry over finite fields presented in [8], this paper introduces a new digital transform, the finite field discrete cosine transform (FFDCT). The FFDCT is defined for signals over GF($p$), $p \equiv 3$ (*mod* 4) and its spectrum has components over GF($p$).

## 2. Background and Preliminaries

2.1 Fundamentals on finite field complex numbers

<u>Definition 1</u>. The set of Gaussian integers over GF($p$) is the set GI($p$) = {$a + jb$, $a, b \in$ GF($p$)}, where $p$ is a prime such that $j^2 = -1$ is a quadratic nonresidue over GF($p$). Only primes of the type $p \equiv 3$ (*mod* 4) meet such a requirement [14]. ❑

The extension field GF($p^2$) is isomorphic to the "complex" structure GI($p$) [15]. From the above definition, GI($p$) elements can be represented in the form $a + jb$ and are referred to as complex finite field numbers.

<u>Definition 2</u>. (unimodular set): The elements $\zeta = (a+jb) \in$ GI($p$), such that $a^2+b^2 \equiv 1$ (*mod p*) are referred to as unimodular elements. ❑

Unimodular elements are used in proposition 3 to construct a GF($p$)-valued DCT.

## 2.2 Finite Field Trigonometry

This session describes briefly trigonometric functions over a finite field, which hold many properties similar to those of the standard real-valued trigonometric functions [16]. In what follows, the symbol := means *equal by definition*.

<u>Definition 3</u>. Let $\zeta$ be a nonzero element of GI($p$), where $p \equiv 3$ (*mod* 4). The *k*-trigonometric functions cosine and sine of $\angle(\zeta^i)$ (arc of the element $\zeta^i$) over GI($p$), are

$$\cos_k(\angle \zeta^i) := (2^{-1} \bmod p) \ (\zeta^{ik} + \zeta^{-ik})$$

and

$$\sin_k(\angle \zeta^i) := (2^{-1} \bmod p) \ (\zeta^{ik} - \zeta^{-ik})/j$$

$i, k = 0, 1,..., N$-1, where $\zeta$ has order $N$. ❑

For the sake of simplicity, these are denoted by $\cos_k(i)$ and $\sin_k(i)$. These definitions make sense only if GI($p$) is a field. This is the reason for requiring $p \equiv 3$ (*mod* 4).

Over the field of real numbers, the DCT is defined by the pair

$$C[k] := \sum_{n=0}^{N-1} x[n] \cos\left( \frac{(2n+1)k\pi}{2N} \right),$$

$$x[n] = \sum_{k=0}^{N-1} \beta[k] C[k] \cos\left( \frac{(2n+1)k\pi}{2N} \right)$$

where $\beta[k]$ is the weighting function

$$\beta[k] = \begin{cases} \frac{1}{2}, & \text{if } k = 0 \\ 1, & \text{if } k = 1 \end{cases}.$$

The steps leading to the expression for $C[k]$, the DCT of the length $N$ time sequence $x[n]$, involve replicating $x[n]$ and computing its DFT, from which the DCT coefficients can be obtained. Therefore, to construct a length $N$ DCT, a kernel of order 2$N$ is required.

## 3. The Discrete Cosine Transform in a Finite Field

Let $f = (f_i)$ be a length $N$ vector over GF($p$). To define its DCT using k-cosines, in a similar way to the classical DCT, the following lemma is required.

**Lemma 1** (k-cos lemma): If $\zeta \in$ GI($p$) has multiplicative order 2$N$, then

$$A = \sum_{k=1}^{N-1} \cos_k(i) = \begin{cases} N-1, & \text{if } i = 0 \\ -1, & \text{if } i \text{ is even} (\neq 0) \\ 0, & \text{if } i \text{ is odd} \end{cases}$$

Proof: By definition

$$A = \sum_{k=1}^{N-1} \cos_k(i) = \tfrac{1}{2} \sum_{k=1}^{N-1} (\zeta^{ki} + \zeta^{-ki}),$$

so that, clearly, $A = N$-1 if $i = 0$. Otherwise,

$$A = \tfrac{1}{2} [\frac{\zeta^i(\zeta^{i(N-1)} - 1)}{\zeta^i - 1} + \frac{\zeta^{-i}(\zeta^{-i(N-1)} - 1)}{\zeta^{-i} - 1}].$$

Since $\zeta$ has order 2$N$, then $\zeta^N = -1$. Multiplying the second term by $(-\zeta^i)$, yields

$$A = \tfrac{1}{2} [\frac{(-1)^i - \zeta^i}{\zeta^i - 1} + \frac{1 - (-1)^i \zeta^i}{\zeta^{-i} - 1}].$$

Therefore, for $i$ even, $A = \tfrac{1}{2} [\frac{1 - \zeta^i + 1 - \zeta^i}{\zeta^i - 1}] = -1$

and, for $i$ odd,

$$A = \tfrac{1}{2} [\frac{-1 - \zeta^i + 1 + \zeta^i}{\zeta^i - 1}] = 0. \qquad ❑$$

From this k-cos lemma, it is possible to define a new digital transform, the finite field discrete cosine transform (FFDCT).

<u>Definition 4</u>: If $\zeta \in$ GI($p$) has multiplicative order 2$N$, then the finite field discrete cosine transform of the sequence $f = (f_i)$, $i = 0,1,... N$ - 1, $f_i \in$ GF($p$), is the sequence $C = (C_k)$, $k = 0,1,... N$ - 1, $C_k \in$ GI($p$), of elements

$$C_k := \sum_{i=0}^{N-1} 2 f_i \cos_k(\tfrac{2i+1}{2}).$$

The inverse FFDCT is given by theorem 1 below.

**Teorema 1** (The inversion formula): The inverse finite field discrete cosine transform of the sequence $C = (C_k)$, $k = 0,1,.. N$-$1$, $C_k \in GI(p)$, is the sequence $f = (f_i)$, $i = 0,1,.. N$-$1$, $f_i \in GF(p)$, of elements

$$f_i = \frac{1}{N} \sum_{k=0}^{N-1} \beta_k C_k \cos_k(\tfrac{2i+1}{2}),$$

where the weighting function $\beta_k$ is given by

$$\beta_k = \begin{cases} (2^{-1} \bmod p), & \text{if } k = 0 \\ 1, & \text{if } k \neq 0 \end{cases}.$$

Proof: To establish the inversion formula, it is sufficient to show that $g_i = f_i$, $i = 0,1,... N$-$1$, where

$$g_i := \frac{1}{N} \sum_{k=0}^{N-1} \beta_k C_k \cos_k(\tfrac{2i+1}{2}).$$

From definition 4, one may write

$$g_i = \frac{1}{N} \sum_{k=0}^{N-1} \beta_k [\sum_{r=0}^{N-1} 2 f_r \cos_k(\tfrac{2r+1}{2})] \cos_k(\tfrac{2i+1}{2}),$$

which is the same as

$$g_i = \frac{2}{N} \sum_{r=0}^{N-1} f_r [\sum_{k=0}^{N-1} \beta_k \cos_k(\tfrac{2r+1}{2}) \cos_k(\tfrac{2i+1}{2})].$$

Using the addition of arcs formula [16]

$$\cos_k(a \pm b) = \cos_k(a)\cos_k(b) \mp \sen_k(a)\sen_k(b),$$

leads to

$$g_i = \frac{2}{N} \sum_{r=0}^{N-1} f_r [\tfrac{1}{2} + \tfrac{1}{2} \sum_{k=1}^{N-1} [\cos_k(r+i+1)] + \tfrac{1}{2} \sum_{k=1}^{N-1} \cos_k(r-i)]].$$

From the k-cos lemma and observing that $(r+i+1)$ is even whenever $(r-i)$ is odd and vice-versa, the evaluation of the above expression requires the consideration of three distinct situations:

i) If $r+i+1 = 0$, then $r = -i - 1$, which implies $f_r = 0$. Therefore, in this case, $g_i = 0$.

ii) If $r-i = 0$, then $r = i$. In this case
$$g_i = \tfrac{2}{N} f_i [\tfrac{1}{2} + \tfrac{1}{2}(0) + \tfrac{1}{2}(N-1)] = f_i.$$

iii) If both, $r+i+1$ and $r-i$ are different from zero, considering the parity for these terms it is possible to write

$$g_i = \tfrac{2}{N} \sum_{r=0}^{N-1} f_r [\tfrac{1}{2} + \tfrac{1}{2}(0) + \tfrac{1}{2}(-1)] = 0,$$

so that $g_i = f_i$, $i = 0,1,... N$-$1$. ❏

To conclude this session, the elements of an FFDCT of length 8 over GF(31) are presented in example 1.

**Example 1**: For $p = 31$, the element $\zeta = (7+j13) \in$ GL(31) has order $(p+1)/2 = 16$. The FFDCT of length $(p+1)/4 = 8$ of the sequence $f = (1, 2, 3, 4, 5, 6, 7, 8)$ is the sequence $C = (10, 20, 0, 17, 0, 12, 0, 5)$. The transform matrix $\{2\cos_k(\tfrac{2i+1}{2})\}$, $i, k = 0,1,..,7$, is

$$M_{k,i} = \begin{bmatrix} 2 & 2 & 2 & 2 & 2 & 2 & 2 & 2 \\ 27 & 10 & 20 & 22 & 9 & 11 & 21 & 4 \\ 14 & 5 & 26 & 17 & 17 & 26 & 5 & 14 \\ 10 & 9 & 4 & 11 & 20 & 27 & 22 & 21 \\ 8 & 23 & 23 & 8 & 8 & 23 & 23 & 8 \\ 20 & 43 & 22 & 10 & 21 & 9 & 27 & 11 \\ 5 & 17 & 14 & 26 & 26 & 14 & 17 & 5 \\ 22 & 11 & 10 & 4 & 27 & 21 & 20 & 9 \end{bmatrix}.$$

The inverse matrix, which is equal to $\{\tfrac{\beta_k}{N}\cos_k(\tfrac{2i+1}{2})\}$, $i, k = 0, 1,..., 7$, is given by

$$M^{-1}{}_{k,i} = \begin{bmatrix} 2 & 23 & 28 & 20 & 16 & 9 & 10 & 13 \\ 2 & 20 & 10 & 18 & 15 & 8 & 3 & 22 \\ 2 & 9 & 21 & 8 & 15 & 13 & 28 & 20 \\ 2 & 13 & 3 & 22 & 16 & 20 & 21 & 8 \\ 2 & 18 & 3 & 9 & 16 & 11 & 21 & 23 \\ 2 & 22 & 21 & 23 & 15 & 18 & 28 & 11 \\ 2 & 11 & 10 & 13 & 15 & 23 & 3 & 9 \\ 2 & 8 & 28 & 11 & 16 & 22 & 10 & 18 \end{bmatrix}.$$
❏

It is interesting to observe, at this point, that, due to the expressions defining the FFDCT pair, there is a simple relation between the direct and inverse transform matrices. In fact, the elements of these matrices are related by

$$m_{i,k}^{-1} = \begin{cases} (2^{-1} \bmod p) \, m_{k,i}, & \text{if } k = 0 \\ m_{k,i}, & \text{if } k \neq 0 \end{cases}.$$

❏

From a practical point of view, an important family of finite field transforms may be obtained from the FFDCT, namely the Mersenne FFDCT. These are defined over GF($p$) when $p=2^s-1$ is a Mersenne prime. The blocklength is $N=2^{s-2}$, which is attractive since that radix-2 fast algorithms can be used in this case.

In general, the transformed vector (DCT spectrum) lies over the extension field GF($p^2$). However, if a unimodular element $\zeta$ is used to define the finite field trigonometry, then it can be shown (proposition 1) that *cos* and *sin* are real functions [17].

Proposition 1. If $\zeta = a + jb$ is unimodular, then $cos_k(i)$ and $sin_k(i) \in$ GF($p$), for any $i, k$. ❏

This is the situation illustrated in example 1. In this case $\zeta$ is unimodular and has a square root $\lambda$ that is also unimodular. This $\lambda$ has order ($p+1$), so that $\zeta$ has order ($p+1$)/2, which implies an FFDCT of length $N = (p+1)/4$. Table 1 below lists the parameters of some real FFDCT.

**Table 1**. Parameters for the FFDCT over a few Finite Fields: Ground field, transform blocklength, unimodular element used to define $cos_k(i)$ and its order over the extension field.

| Ground field GF($p$) | Block length $N$ | unimod element $\zeta$ | Extention field GF($p^2$) | Order Ord($\zeta$) |
|---|---|---|---|---|
| 7* | 2 | 2+j2 | GF(49) | 8 |
| 23 | 6 | 4+j10 | GF(529) | 24 |
| 31* | 8 | 2+j11 | GF(961) | 32 |
| 47 | 12 | 4+j19 | GF(2209) | 48 |
| 71 | 18 | 8+j24 | GF(5041) | 72 |
| 79 | 20 | 2+j32 | GF(6241) | 80 |
| 103 | 26 | 2+j10 | GF(10609) | 103 |
| 127* | 32 | 2+j39 | GF(16129) | 128 |
| 151 | 38 | 2+j65 | GF(22801) | 152 |
| 167 | 42 | 4+j73 | GF(27889) | 168 |
| 191 | 48 | 6+j27 | GF(36481) | 192 |
| 199 | 50 | 2+j14 | GF(39601) | 200 |

* Mersenne FFDCT.

## 4. Conclusions and Suggestions

In this paper the discrete cosine transform in a finite field GF(p) was introduced. The FFDCT uses the k-trigonometric $cos_k(.)$ function over GF(p) as kernel. An important lemma concerning such functions was given, from which the inversion formula was established. The length of the transform is a divisor of $p+1$, where $p \equiv 3$ (*mod* 4). If $p$ is a Mersenne prime, the corresponding transform is called Mersenne FFDCT. In this case the transform blocklength is a power of two. Unimodular elements were selected to guarantee real values for the k-trigonometric functions, thus producing real (GF($p$)-valued) transforms.

Since there exists many definitions for the classical DCT, it is interesting to investigate alternative definitions for the FFDCT as well. Further Transforms such as discrete sine transform and 2-D FFDCT could also be defined (DST, 2D-FFDCT). Possible aplications of the FFDCT for multiplex, CDMA schemes and image processing are currently under investigation.